\documentclass[graybox]{svmult}
\usepackage{type1cm} 
\usepackage{makeidx} 
\usepackage{multicol}
\usepackage[bottom]{footmisc}
\usepackage{graphicx}
\usepackage{caption}
\usepackage{subcaption}
\usepackage{todonotes}
\usepackage{xcolor}
\usepackage{xspace}
\usepackage{cleveref}
\usepackage{algorithm}
\usepackage[noend]{algpseudocode}
\usepackage{authblk}
\usepackage{tcolorbox} 

\newcommand\R[1]{{\color{black}#1}}
\newcommand\B[1]{{\color{black}#1}}

\begin{document}

\title*{\B{Teaching Mining Software Repositories}}

\author{Zadia Codabux, Fatemeh Fard, Roberto Verdecchia, Fabio Palomba, Dario Di Nucci, Gilberto  Recupito}

\institute{Zadia Codabux 
\at University of Saskatchewan, Canada, \email{zadiacodabux@ieee.org}
\and Fatemeh Fard 
\at University of British Columbia, Canada, \email{fatemeh.fard@ubc.ca}
\and Roberto Verdecchia 
\at University of Florence, Italy, \email{roberto.verdecchia@unifi.it}
\and Fabio Palomba 
\at University of Salerno, Italy, \email{fpalomba@unisa.it}
\and Dario Di Nucci 
\at University of Salerno, Italy, \email{ddinucci@unisa.it}
\and Gilberto Recupito 
\at University of Salerno, Italy, \email{grecupito@unisa.it}}

\maketitle

\noindent \textbf{Abstract}\\

\B{Mining Software Repositories (MSR) has become a popular research area recently. MSR analyzes different sources of data, such as version control systems, code repositories, defect tracking systems, archived communication, deployment logs, and so on, to uncover interesting and actionable insights from the data for improved software development, maintenance, and evolution. This chapter provides an overview of MSR and how to conduct an MSR study, including setting up a study, formulating research goals and questions, identifying repositories, extracting and cleaning the data, performing data analysis and synthesis, and discussing MSR study limitations. Furthermore, the chapter discusses MSR as part of a mixed method study, how to mine data ethically, and gives an overview of recent trends in MSR as well as reflects on the future. 
As a teaching aid, the chapter provides tips for educators, exercises for students at all levels, and a list of repositories that can be used as a starting point for an MSR study. \\ 
 }

\section{Introduction}
\R{Embedded within the intricate software development landscape lies Mining Software Repositories (MSR), an academic discipline dedicated to delving into the expansive dataset within software repositories.
Software Repositories are among the most common sources of data that enable the analysis of software properties. 
\R{Additionally, the promising success that this type of study had in the research community allowed to extend the mining data process beyond the software repositories, encompassing various essential aspects such as software changes, collaborative efforts, socio-technical aspects~\cite{gousios2008measuring,mens2016ecosystemic}, and the process dynamics~\cite{poncin2011process}.}
Combining these aspects serves to obtain the source of key information detailing the transformation of a software application from its preliminary stages to well-defined and sophisticated software.}

During the iterative enhancement of software development, developers wield powerful tools that meticulously track software information. These tools, including robust version control systems like Git and Subversion (SVN) and efficient issue-tracking tools like JIRA and Bugzilla, facilitate collaboration and provide an exhaustive lineage of code changes. This comprehensive history, resulting from the union of these tools, allows developers to gain profound insights into all key aspects of the software evolution project, including software modifications, issue resolutions, bug fixes, and strategic refactoring initiatives, which are contained in the software repository~\cite{kagdi2007survey}.
The definition of this dynamic and extensive source of information lays the basis for introducing MSR. MSR is the field that exploits the information contained in software repositories to conduct thorough investigation and analysis, unveiling aspects of the software development process ~\cite{hassan2008road}.

MSR studies in the software engineering research field demonstrated the underlying capabilities of repository analysis. Several studies investigate the possibility of building prediction models for software quality issues. Prediction models could be built from issue tracking labels to detect bugs ~\cite{marcus2004information,rao2011retrieval}. Moreover, the extreme amount of data possible to extract from source code repositories allows us to build complex and useful models with the use of Large Language Models, enabling us to build useful tools that perform programming tasks related to code generation, summarization, and analysis, also producing code that is not complex and highly maintainable (e.g., Copilot\footnote{Github Copilot: https://github.com/features/copilot/})~\cite{Nguyen2022CopilotMSR}.
\R{This chapter is intended to \textbf{target educators at MSc and Ph.D. levels}, providing them with a comprehensive understanding of MSR and its applications in software engineering research so that they may be equipped to teach this method. More specifically, the \textbf{learning objectives} include:}

\begin{enumerate}
    \item \R{Understand \textbf{how to set up a mining software repository study}, uncovering the methodological design steps required to (i) define the objectives of an MSR study; (ii) identify suitable data sources and data cleaning methods to address the objectives; (iii) select the most appropriate code analysis instruments to extract relevant pieces of information useful to the analysis; and (iv) analyze and synthesize the data gathered throughout the mining process;}

    \item \R{Realize \textbf{what are the typical threats to the validity involving mining software repository studies}, overviewing how multiple issues concerned with the mining and analysis of data sources might bias the interpretation of the results;}

    \item \R{Identify \textbf{how to combine mining software repository studies with additional research methods}, discussing the flexibility of MSR research in addressing complex research objectives;}

    \item \R{Indicate \textbf{what are the ethical considerations concerned with mining software repository studies}, uncovering the possible limitations of this research method, other than the responsibilities that the future generation of researchers must necessarily take into account when designing similar studies.}
\end{enumerate}

To ease the reader's understanding of the concepts in the chapter, \R{we will focus on specific cases regarding extracting data from the source code of software repositories.} Specifically, we will often refer to an exemplar case, which effectively used the methodologies described in the chapter. Among the vast array of valuable MSR examples in literature, we decided to focus on the work by Kamei et al.~\cite{Kamei2013JITQA}, published in IEEE Transactions on Software Engineering in 2012. 

The study explored a well-known and well-established research theme in MSR research, namely \emph{defect prediction}. This is a technique used to identify defect-prone files or packages throughout software development. More particularly, the study introduced the term \emph{``Just-In-Time Quality Assurance''} and aimed at identifying defect-prone software changes while the developers commit their changes onto a shared repository. From a methodological standpoint, the study heavily relied on MSR instruments, as it was required to feed machine-learning models with features mined from software repositories. These features encompassed multiple properties that might be extracted from version control systems, such as product, process, and developer-oriented metrics, hence representing an ideal example to make the concepts in this chapter more practical. In addition, the study carefully designed the experimental procedures, touching on various data selection, cleaning, and analysis processes that are elaborated upon in this chapter. 

\R{Exemplary \textbf{teaching strategies} that might be relevant for teaching this chapter include, but are not limited, to the following:}

\begin{itemize}
    \item \R{\emph{Interactive discussions}. Educators may engage students in active discussions about the importance of MSR in software engineering research, its potential applications, and its limitations. Through these discussions, we envision students to acquire awareness of when and how to use MSR instruments to empower their analyses;}

    \smallskip 
    \item \R{\emph{Case studies}. Educators may use our chapter to present real-world examples of MSR research, such as the study by Kamei et al.~\cite{Kamei2013JITQA}, to illustrate key concepts and methodologies, making students aware of how the design choices taken in exemplary cases impacted the type of analysis and results of MSR studies;}

    \smallskip
    \item \R{\emph{Hands-on exercises}. Educators may provide students with hands-on experience in conducting MSR studies, including data collection, cleaning, and analysis, using relevant tools and datasets. Educators may require students to design an MSR study, producing reports that detail the whole set of activities and the rationale thereof required to address specific research objectives;}

    \smallskip
    \item \R{\emph{Group projects}. Educators may assign group projects where students can apply MSR techniques to analyze software repositories, like those reported at the end of this chapter, and address specific research questions or challenges. Using group projects, students might jointly design data collection, cleaning, and analysis steps for specific research objectives, having the chance to interact with other students, hence possibly increasing the collective awareness of the most appropriate methodological choices to take, even considering ethical limitations. }
\end{itemize}

\section{Setting up a Source Code Repository Mining Study}

\begin{figure}
    \centering
    \includegraphics[width=\linewidth]{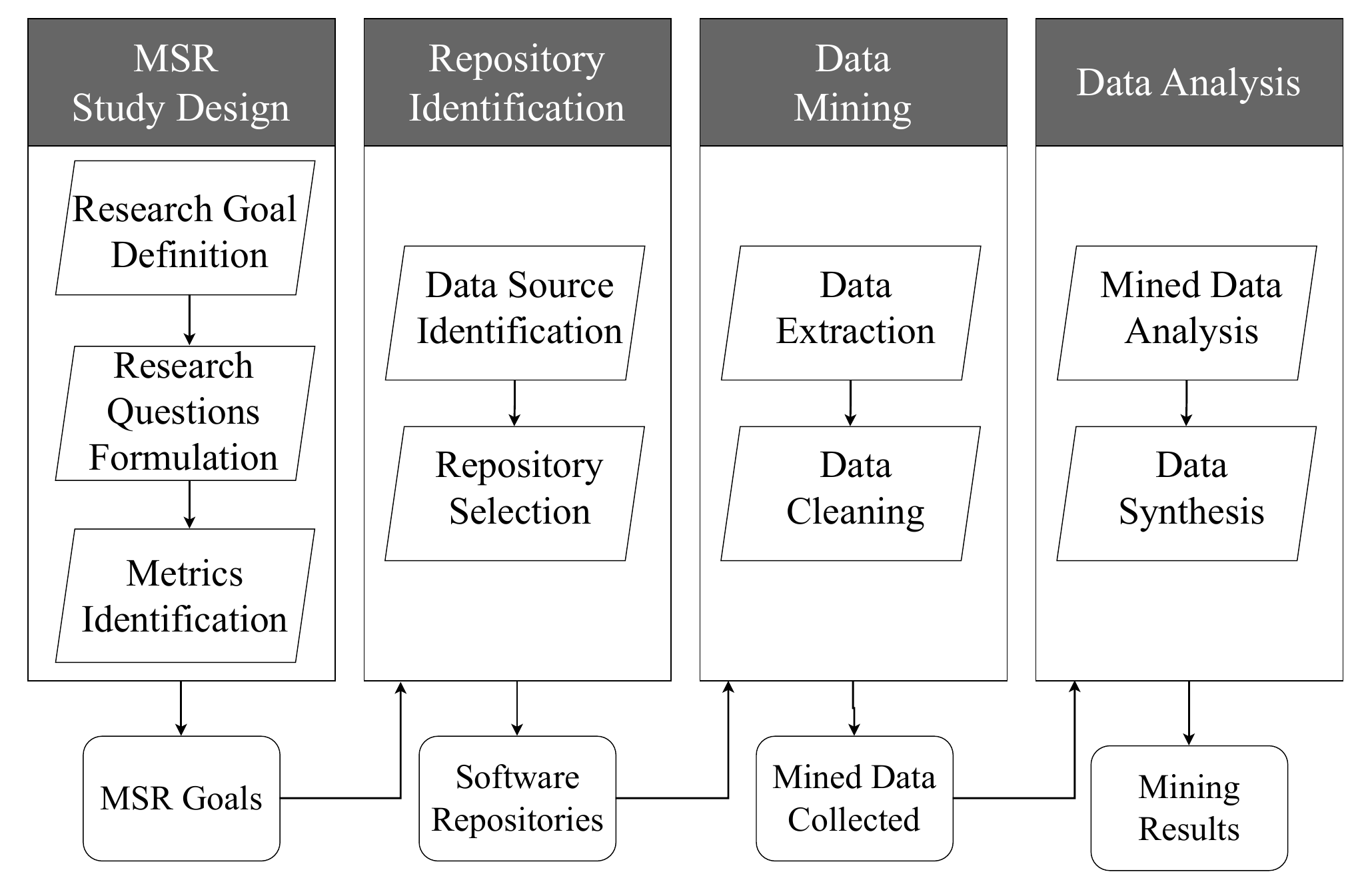}
    \caption{\R{Set of Activities Involved in the MSR Study Process.}}
    \label{fig:msr_process}
\end{figure}

This section will overview the MSR process, including the study design, data extraction, and analysis, as illustrated in Figure \ref{fig:msr_process}. We will discuss the considerations for data extraction (e.g., API vs. Package) and storage. 
\R{Since these steps are typically included to guarantee a systematic mining process, the definition of it might vary depending strictly on the goal of the study that an MSR researcher aims to design.}
Regarding data analysis, we will focus on aspects that should be considered and errors that could occur about the storage, dataset size, types of analysis, emergent problems, and cases of data contradicting Research Questions (RQs) \R{or formulated hypotheses}.\\

\subsection{MSR Study Design}
\label{msr_process:gqm}
The first step to designing an MSR study should be systematically defining the research goal. To do so, the widely adopted Goal-Question-Metric (GQM) paradigm of Basili~\cite{basili1994goal} can be used. Designing the objective of a study \textit{via} the GQM paradigm is composed of three main steps, namely (i) the formulation of the goal of the research, (ii) the design of the RQs that need to be answered to achieve the goal, and (iii) the definition of the metrics necessary to answer the goal.

\subsubsection{Research goal formulation}
\label{msr_process:goal}
The formulation of the research goal of a mining study through the GQM paradigm follows a systematic structure. More specifically, the goal is designed by adopting the following template:

\begin{quote}
    \textbf{Analyze} \textit{the experimental object(s) of the study, e.g., repository source code or GitHub issues}\\
    \textbf{For the purpose of} \textit{the focus of the study, e.g., identifying common bug types or understanding developer behavior} \\
    \textbf{With respect to} \textit{the quality focus, e.g., functional suitability or performance}\\
    \textbf{From the viewpoint of} \textit{the intended reader of the study, e.g., developers or researchers}\\
    \textbf{In the context of} \textit{the context considered, e.g., the Apache ecosystem or Android apps}.
\end{quote}

On the one hand, formulating the goal according to the template allows one to take the time to reason on the core objectives of the research and start reflecting on the research method to achieve such a goal. On the other hand, reporting the goal by following the GQM template allows one to systematically document and swiftly communicate the research intention to those not conducting the study. 

As an example of how a research goal can be formulated according to the GQM template, let us consider the study of Kamei et al.~\cite{Kamei2013JITQA}. In the case of this study, a possible formulation of the research goal could be:

\begin{quote}
    \textbf{Analyze} \textit{defect prediction models}\\
    \textbf{for the purpose of} \textit{defect prediction} \\
    \textbf{with respect to} \textit{risky changes} \\
    \textbf{from the point of view of} \textit{software developers and reviewers} \\
    \textbf{in the context of} \textit{open source and commercial projects from multiple domains}.
\end{quote}

Note that, while introducing some systematicity, a single research goal could be correctly formulated in many ways while following the GQM template. Following Occam's razor, the most simple and informative solution should be preferred among different options. Ideally, the research goal is jointly discussed by all people partaking in the mining process. It serves as a collective moment of reflection on the goal and intent of the mining procedure.

\begin{tcolorbox}[colback=white!5!white,colframe=blue!75!black]
  \R{\textbf{A Do for Educators.} When formulating research goals, among different options, follow Occam's razor; your students should prefer the most simple and informative solution.}
\end{tcolorbox}

\subsubsection{Research Questions Definition}
Once the study's goal is defined according to the GQM template (see previous section), the RQs the mining process aims to answer can be formulated.
Formulating the RQs based on the research goal allows for further design, with a systematic step-by-step process, of the mining to be executed.
RQs should be directly derived from the research goal formulated in the previous step. Once an answer to each RQ is provided, it is possible to assess the extent to which the research goal is achieved (and, in negative cases, further enhance the research process by adding further mining processes or RQs). 
If needed, during a preliminary repository mining design phase, RQs can be adapted to fit the refined research goal and envisioned repository mining process.

\B{As noted in a recent work by Storey et al.~\cite{storey2024disruptive} when conducting MSR studies, RQs are often formulated in a rushed manner. To avoid this, a deliberate approach to infer the most suited RQs is recommended. This involves considering the potential impact of the research, drawing inspiration from various sources, and intersecting these with possible phenomena and concepts related to those phenomena. The next step is to brainstorm RQs. This is a creative process where no idea is initially dismissed. The aim is to generate a wide range of potential research questions that could be studied.
Once a comprehensive list of potential research questions has been generated, the final step is to select specific research questions for further study. This selection should be justified based on the relevance of the question to the research objectives, the feasibility of answering the question, and the potential contribution of the findings to the field of study.}

As a rule of thumb, a mining process is steered by two or three RQs. Having more than four or five RQs usually suggests that the research goal is not well-defined or that the RQs are too low-level.
By taking into account the exemplary study of Kamei et al.~\cite{Kamei2013JITQA}, we can see how their research goal is decomposed into three different research questions, namely:

\begin{quote}
$RQ_1:$ \textit{How well can we predict defect-inducing changes?}\\
$RQ_2:$ \textit{Does prioritizing changes based on predicted risk reduce review effort?}\\
$RQ_3:$ \textit{What are the major characteristics of defect-inducing changes?}
\end{quote}

By comparing the RQs of Kamei et al.~\cite{Kamei2013JITQA} with the research goal formulated in Section~\ref{msr_process:goal}, we can observe that $RQ_1$ is the one more closely related to the goal. At the same time, $RQ_2$ and $RQ_3$ are utilized to build upon the research goal and provide further complementary notions on the considered topic. While slightly striving away from the standard application of the GQM approach, such a technique can be used to build upon and strengthen the results of the mining process set by the research goal.

\R{As potential inspiration, examples of further RQs of an MSR study could be:\\
$RQ:$ \textit{Can Large Language Models be used to identify bugs?} \\
$RQ:$ \textit{Is there a relation between technical debt and software energy efficiency?}\\
$RQ:$ \textit{How does software quality evolve in microservice architectures?}\\
$RQ:$ \textit{Which code smells are more frequent in AI-centric software projects?}\\
$RQ:$ \textit{What are the most common causes of test flakiness?}\\
}

From a documentation point of view, once the mining results are collected and analyzed, it is considered a good practice to explicitly answer each RQ (and possible sub-RQs if present) in the mining process documentation. Explicit answers to the RQs are usually documented either in the Results section, Discussion section, or a section dedicated entirely to answering the RQs. 

\begin{tcolorbox}[colback=white!5!white,colframe=blue!75!black]
  \R{\textbf{A Do for Educators.} A mining process is usually steered by two or three RQs. Usually, having more RQs suggests that the research goal needs to be better defined or that the RQs need to be more high-level.}
\end{tcolorbox}

\subsubsection{Metrics identification}
As the last step of the GQM approach, each RQ is mapped to a specific set of metrics that do not overlap with those used to answer other RQs. However, this is not always the case. For example, the number of GitHub issues could be used to answer both RQs regarding developer behavior and recurrent development impediments. The selection of metrics strictly depends on the specific RQs at hand, which can be qualitative, quantitative, or a mix of both. 
\R{Selected metrics can be either atomic, e.g., lines of code or number of defects, or composite, e.g., defects per line of code. It is, therefore, possible to compose the same set of atomic metrics in different manners to answer different RQs.}

\begin{tcolorbox}[colback=white!5!white,colframe=blue!75!black]
  \R{\textbf{A Do for Educators.} When documenting the metrics used to answer the RQs, students should report their definitions and mappings to the RQs, supporting the metric definition with a sound reference or an unequivocal definition and measurement procedure description.}
\end{tcolorbox}

\begin{tcolorbox}[colback=white!5!white,colframe=blue!75!black]
  \R{\textbf{Teaching Exercises.} Through the GQM approach, define a goal, formulate three RQs, and identify five metrics (mapped to the RQs) to investigate which development factors affect bug-proneness. }
\end{tcolorbox}

\subsection{Identifying High-Quality Source Code Repositories}

\subsubsection{Data Sources} 
The MSR field collects and analyzes rich data from different types of repositories. The repositories can be of static and dynamic nature and can be classified as follows~\cite{hassan2008road}:

\begin{itemize}
    \item Historical repositories, including 
    (i) bug or defect repositories, such as Jira and Bugzilla, which help track and manage defects of a software system,
    (ii) archived communications such as mailing lists, emails, and chat messages, which are a source of discussions regarding multiple aspects of a software system, and
    (iii) source control repositories that track all the changes made to software system artifacts (e.g., source code, Pull Requests (PRs), commit messages, documentation). Git is one of the most popular source control repositories. 
    \item Source Code repositories control and manage software projects. Typical features include source code hosting, bug tracking, documentation facilities, mailing lists, and forums. Notable source code repositories include Sourceforge, GitHub, and Bitbucket. 
    \item Runtime repositories such as deployment logs that track the information and actions related to the deployment of a software system. Deployment logs are essential to oversee the configurations and steps of deployment instances. They usually include timestamps, error messages, and configuration settings. 
\end{itemize}

\R{For this chapter, we focused on source code repository mining to align with the running exemplar.}

\subsubsection{Source Code Repositories Selection Criteria}

Code repositories have multiple characteristics~\cite{kalliamvakou2014promises} that can help in the decision-making process for selection. These characteristics can be used as inclusion and exclusion criteria for narrowing down the candidate repositories. 

\begin{itemize}
    \item Programming language(s), e.g., Java for the Apache Ecosystem projects. 
    \item Size and complexity of the repository, e.g., number of lines of code (LOC) or number of commits
    \item Domain, e.g., gaming, visualization, database, parsers, testing. 
    \item Active or inactive repository, e.g., were there recent (during the last 6 months) commit activities on the repository? When was the last commit?
    \item Types of repositories, i.e., base (not forked) or forked repositories. 
    \item Purpose of the repository. Not all repositories are software repositories. Some are used for experimentation, website hosting, academic, and personal (not involving collaboration) projects.
    \item Location of the repository. Some projects are hosted on multiple platforms. 
    \item Popularity of the repository, e.g., number of stars and contributors. 
\end{itemize}

\subsubsection{Considerations when Selecting Data Sources}

Several considerations must be undertaken to mitigate potential threats when assessing source code repositories for research purposes.
The repositories encompass diverse entities beyond software projects, serving as repositories for free web storage, online books, or repositories housing projects with other characteristics. 
Numerous repositories exhibit transient traits, being short-lived, inactive, or dedicated to assignments, student endeavors, educational objectives, personal use, or archival purposes~\cite{kalliamvakou2014promises}. 
Thus, if the research question is related to software development, these projects should be removed from the dataset. 
Furthermore, it is crucial to prioritize repositories that house engineered software projects~\cite{munaiah2017curating}.
Consequently, in studies related to software development, it becomes imperative to purify the dataset for such repositories. 
\R{Munaiah et al.~\cite{munaiah2017curating} propose a framework to filter the engineered software projects from GitHub.}
Additionally, many projects have few commits, and not all use pull requests, resulting in skewed or imbalanced data. 
The process of commits in pull requests and for code reviews depends on GitHub's practice of recording the commits, which should also be studied carefully.

\R{A strategy to mitigate such potential threats} involves refraining from considering repositories that include many non-registered users as committers. 
Furthermore, projects explicitly identifying themselves as mirrors in their descriptions should be carefully vetted or excluded.
\B{For an MSR study, we recommend having a list of inclusion and exclusion criteria based on study goals and RQs. Manual exploration can be conducted on sample repositories to assess whether any changes to the selection criteria are required to ensure a high-quality dataset is collected. } 
 
When selecting the repositories, researchers should note that selecting and ranking the repositories based on the number of stars might favor active marketing strategies and not the well-established software engineering practices~\cite{BORGES2018112}. Therefore, checking that the repositories are not starred in a short period due to social media activity is essential.  
It is also important to note that the number of stars is not strongly correlated with contributors, forks, commits, and the repository's age. Thus, relying only on the number of stars can threaten the validity of the collected data.

Another consideration is the correctness of the heuristics and key terms used for selecting the repositories. These heuristics should be scrutinized and documented~\cite{hassan2008road}. The criteria for choosing the data sources may lead to a noisy collection of repositories and, therefore, a skewed dataset. Careful investigations reveal whether this skewness is related to the inaccurate metrics for selecting repositories or is due to the nature of the data. 

Understanding the limitations of repository data is another consideration, as the repository data cannot lead to causal conclusions and only can show correlations~\cite{hassan2008road}. Moreover, the active projects might not include all their development activities in GitHub. So, other resources should be considered.

\R{In general, MSR findings must be considered in the context in which the \B{studies} are performed, which is crucial to revealing the actual cause of particular conclusions. Indeed, such findings may not generalize across projects, and repository use could vary between projects. So, researchers should closely examine socio-technical aspects~\cite{hoda2021socio} to better understand the use of repositories before reaching conclusions.

When presenting the results of MSR research, the limitations of repository data should be thoroughly examined and communicated. This practice is essential to prevent misinterpretations and ensure research integrity.}

\begin{tcolorbox}[colback=white!5!white,colframe=blue!75!black]
  \R{\textbf{A Do for Educators.} Students should carefully select the sources before analyzing them. When selecting data sources, the above-stated considerations allow for effective mining, which is essential to answer the RQs, prevent misinterpretations, and ensure research integrity.}
\end{tcolorbox}

\begin{tcolorbox}[colback=white!5!white,colframe=blue!75!black]
  \R{\textbf{Teaching Exercises.} A typical exercise would be to select three GitHub projects by focusing on three inclusion and exclusion criteria each, e.g., projects written in different programming languages, of different domains, and active projects with at least 200 commits over the last year as inclusion criteria. Exclusion criteria could include academic and personal projects and projects hosted on multiple platforms. Once the projects have been shortlisted, the issue-tracking data can be mined for further analysis.}
\end{tcolorbox}

\subsection{Data Extraction}

Data extraction starts once a source code repository is selected and depends on the goals of the mining campaign.
\R{In studies concerning predictive analytics, this phase entails mining data concerning independent and dependent variables. The goal is to extract data concerning the former variable able to accurately predict the former variable in the future.}
For example, let us consider just-in-time defect prediction~\cite{Kamei2013JITQA}, which aims to predict defective commits given their features
\R{and involves mining commit data to extract valuable information to forecast such defective commits.}
Such commits can be analyzed sequentially, or specific commits can be analyzed in a given time window.

The dependent variable we will need to extract is the failure proneness of our commits.
In contrast, the independent variables are the characteristics that should lead to such failure.
It is essential to consider that the described concepts are generalizable to predict defects~\cite{Kamei2013JITQA}, code smells~\cite{azeem2019machine}, and security vulnerabilities~\cite{yamaguchi2011vulnerability} concerning not only traditional code but also Infrastructure-as-Code~\cite{dalla2021within}.

In just-in-time defect prediction, the first step is identifying \textit{defect-fixing} commits.
To achieve this goal, we need to collect the issues closed and related to bugs (e.g., with labels \texttt{bug} and \texttt{bugfix}).
Source code repositories like GitHub provide issue trackers that link issue reports and bug-fixing commits.
A commit message is tagged as fixing defect if it matches a regular expression like the following:

\begin{center}
  (\textit{bug}\textbar \textit{fix}\textbar\textit{error}\textbar\textit{crash}\textbar\textit{problem}\textbar\textit{fail}\textbar\textit{defect}\textbar\textit{patch})
\end{center}

Only the commits that modify at least one source code file are kept.

The second step is to identify the \textit{fixed files}, which are the files modified by \textit{defect-fixing} commits, and their related \textit{defect-inducing} commits.

The commits from the most recent to the oldest are analyzed.
For each \textit{fixed file}, the \textit{SZZ} algorithm~\cite{sliwerski2005changes} automatically identifies the \textit{oldest} commit that modified the lines of code involved in the fix.
It is worth noting that SZZ has evolved over the years, and several variants are available~\cite{rosa2021evaluating}.
Once the defect-inducing commit is found, all commits between the defect-inducing commit (inclusive) and the defect-fixing commit (exclusive) are labeled \textit{failure-prone}. 

After the failure-proneness of the code components has been determined, metrics able to predict such failure can be mined.
In the example, Kamei et al.~\cite{Kamei2013JITQA} mined 14 metrics grouped into five dimensions, i.e., diffusion, size, purpose, history, and experience.
It is essential to consider that although some metrics can predict several phenomena (e.g., the file size could lead to defects and bugs), different metrics apply in different contexts.
Structural metrics~\cite{chidamber1994metrics} focus on structural properties extracted through source code analysis. 
Delta metrics~\cite{di2019delta} capture the amount of change in a file between two successive releases. 
Process metrics~\cite{moser2008comparative} consider aspects concerning the development process rather than the code itself.

In just-in-time defect prediction, all commits could be used to train the models\footnote{In this scenario, the commit order is particularly relevant~\cite{falessi2020need}.}.
Nevertheless, analyzing all commits could be infeasible in other contexts; therefore, selection strategies should be applied.
For example, only one snapshot for each software release could be analyzed by randomly selecting it or applying another strategy.
For instance, the first, last, or middle snapshot of each release could be analyzed. 

Finally, several tools are available to mine source code repositories and extract valuable metrics.
Among them, PyDriller~\cite{spadini2018pydriller} is a Python framework that helps developers mining software repositories.
PyDriller allows the extraction of information from any Git repository, such as commits, developers, modifications, diffs, and source codes, and quickly exports to CSV files.

\begin{tcolorbox}[colback=white!5!white,colframe=blue!75!black]
  \R{\textbf{A Do for Educators.} There is no silver bullet. Although some metrics can predict several phenomena, different metrics apply in different contexts.}
\end{tcolorbox}

\begin{tcolorbox}[colback=white!5!white,colframe=blue!75!black]
  \R{\textbf{A Do for Educators.} Do not re-invent the wheel. Several tools are available to mine source code repositories and extract useful metrics. Before developing a brand-new tool for mining software repositories, explore existing ones.}
\end{tcolorbox}

\subsubsection{Data Cleaning}
\R{Software repositories often contain various types of noise that can skew the results of analyses~\cite{barros2021mining}.}
Therefore, data cleaning is a critical step in preparing data, which includes a range of techniques and procedures to ensure that datasets are accurate, consistent, and reliable.
\R{Several techniques are employed to ensure the adoption of high-quality data before analysis, a crucial practice, particularly in the context of Artificial Intelligence (AI).}
Different methods of handling data greatly depend on the specific context in which it will be used. 

This section is designed to help readers manage data in the context of MSR, emphasizing one of the most critical aspects of software repositories: the commit.
This element can hinder different issues after the data are extracted and can lead to drift in the results of our analysis.
Commits are the puzzle pieces that make up a project's story. Each commit contributes to the bigger picture, but sometimes, some do not fit perfectly, leaving gaps or distorting the intended image. Addressing these problematic commits is essential to maintaining the integrity of the entire project and ensuring the accuracy of our analyses.

When mining data from commits, the first issue concerns extracting data from \textit{tangled commits}~\cite{herzig2013impact}.
This type of commit contains multiple pieces of information related to different changes, caused often by the introduction of \textit{multiple-fixing change commits}~\cite{nguyen2013filtering}. 
These changes during the development phase may not cause major issues. However, they can create problems while analyzing the corresponding version archive by introducing inaccuracies. For example, if a complex change is made to fix a bug, all the files associated with it may be mistakenly labeled as defective in the historical context.
When treating these types of commits, it is suggested to understand the changes inside a commit and consider it separately as \textit{change set partitions}~\cite{herzig2016impact}.

An additional critical issue that arises during the data collection process is the inclusion of \textit{merge commits}, which has been highlighted as a nuanced issue by Kovalenko et al.~\cite{kovalenko2018MSRMergecommits}. \R{This issue requires an additional effort since many merge operations may not be labeled as such, leading to inspecting the content of the changes manually to identify it \cite{kalliamvakou2014promises}.}
\R{Similar to \textit{merge commits}, it is necessary to consider other types of changes and commits that can mislead the analysis. 
In this set, we can also find \textit{quickly remedy commits} (i.e., commits aimed at implementing changes omitted in the previous commit)~\cite{wen2020quickcommits}.
The decision to include or exclude these commits can significantly impact the complexity of the dataset and the richness of the information obtained. 
Wen et al. discovered that excluding this type of commit allows us to avoid introducing a significant amount of noise in MSR studies ~\cite{wen2022quick}.}
Incorporating all commits, including those outside the main branches, can elevate the complexity of the analysis. However, it can also enrich the dataset by adding important information to the software history and providing a more comprehensive view of collaborative efforts, branching strategies, and concurrent development threads.

On the other hand, limiting the analysis to the main branch can simplify the extraction of essential information. However, it may also lead to missing valuable insights into the other branches. Therefore, the inclusion or exclusion of these commits presents a delicate balance, and MSR researchers should carefully consider the trade-offs involved. As such, it is crucial to weigh the benefits and drawbacks of each approach, keeping in mind the research goals and the nature of the data under investigation.

When determining whether to include or exclude commits based on the goals of the MSR study, it becomes evident that commits containing information not aligned with the study's objectives are considered noise.
\textit{Noisy commits} in software development and version control systems refer to changesets or commits deemed extraneous to the primary focus of analysis. These typically involve minor modifications like fixing typos, adjusting task names, or resolving lint warnings \cite{liu2018noisycommits}.
In MSR studies, researchers often encounter noisy commits and filter them out to improve the signal-to-noise ratio, allowing for a more targeted analysis of substantive changes. Dalla Palma et al.~\cite{dalla2021within}, for instance, exemplified this approach by excluding commits that addressed typos, task names, and lint warnings, prioritizing more impactful contributions to the codebase. 
\R{Additionally, an MSR researcher should consider including or excluding bot commits~\cite{tapajit2020botcommits}. 
Detecting and considering the exclusion of this type of commit can significantly differ on several aspects involved in MSR analysis, including community-related aspects~\cite{dey2020detecting}.
This commit type is detectable by observing the commit message and the list of changes the bot will perform.}

To conclude, an MSR researcher must carefully navigate the decisions surrounding commit inclusion or exclusion, acknowledging the significance of commits in shaping the quality of data analysis. By aligning these decisions with the specific goals of the study, researchers ensure that the dataset remains focused on substantive contributions, enhancing the precision and relevance of their findings.
\R{
Therefore, an MSR researcher should take the following steps to obtain the optimal dataset:
\begin{itemize}
    \item Identifying the criteria for selecting the key commits that should be included in the study.
    \item Consider including or excluding tangled, merge, and quick-remedy commits.
    \item Based on the defined criteria, identify and exclude the noisy commits in the extracted data.
\end{itemize}
}
\begin{tcolorbox}[colback=white!5!white,colframe=blue!75!black]
  \R{\textbf{A Do for Educators.} Students should carefully clean data. Each piece of data is the puzzle piece that makes up the story of an MSR campaign. They all contribute to the bigger picture, but sometimes, some do not fit perfectly, leaving gaps or distorting the intended image.}
\end{tcolorbox}

\begin{tcolorbox}[colback=white!5!white,colframe=blue!75!black]
  \R{\textbf{Teaching Exercises.} A practical exercise for understanding the data extraction phase involves students applying algorithms from the study of Kamei et al.~\cite{Kamei2013JITQA} to identify bug-inducing and fixing commits in repositories. Starting with chosen repositories, students identify commits that fix defects. During this process, they meticulously clean the data by removing any tangled and merge commits. Finally, students extract the commit message and the set of files modified by each defect-fixing commit.}
\end{tcolorbox}
    
\subsection{Code Analysis}
Source code analysis is extracting information about a program from its source code or artifacts (e.g., from Java byte code or execution traces) generated from the source code using automatic tools~\cite{binkley2007source}. Source code analysis can be textual, static, dynamic, and historical~\cite{dit2013feature}. We focus on static code analysis. 

\subsubsection{Static Code Analysis}
Source code is a crucial artifact that can be analyzed to reveal information. A technique to analyze source code is static code analysis (also known as structural analysis) - the process of checking the source code of a program for issues without executing the program. Static code analysis is helpful in 
(i) ensuring that code adheres to rules about good coding practices, 
(ii) finding defects and security issues and 
(iii) identifying code smells and technical debt instances. Static analysis can be applied during the early stages of software implementation for early fault detection since the code does not need to be fully functional or executable.  

Source code can be analyzed at different levels of granularity. The analysis level depends on the study's goal and research questions, which will, in turn, dictate the level of granularity at which the analysis needs to be performed. Generally, analysis can be performed at 
(i) method level - the lowest level of granularity which allows finer-grained analysis, 
(ii) class level - a coarser level of granularity compared to method level but still popular among researchers, 
(iii) file level  - a higher level of granularity and lower level of details when analyzing groups of classes and 
(iv) system level - the highest granularity level and the lowest details level.\\

Tools such as \textbf{Automated Static Analysis Tools (ASATs)} are increasingly used for static analysis. ASATs can be of general purpose (e.g., PMD, SonarQube, Understand, Cppcheck), bug-focused (e.g., SpotBugs, Coverity), and security-focused (e.g., Flawfinder, Fortify). 
\B{Static analyzers differ based on the level of sensitivity regarding precision and analysis time~\cite{emanuelsson2008comparative}. A flow-sensitive analysis is more precise but more time-consuming and considers the order of the statements compared to a flow-insensitive analysis. A path-sensitive analysis considers only valid paths and can be precise but costly, while a path-insensitive analysis considers all execution paths, including infeasible ones. Context-sensitive (also known as inter-procedural) analysis considers the context during the analysis compared to an intra-procedural analysis. The latter is faster but more imprecise compared to an inter-procedural analysis.}

The Open Worldwide Application Security Project (OWASP)\footnote{https://owasp.org/www-community/Source\_Code\_Analysis\_Tools} provides a comprehensive list of security static analyzers. Some commonly used static analyzers are presented next.\\

PMD\footnote{https://pmd.github.io/} is a general-purpose static analyzer that supports 16 languages but primarily focuses on Java. For instance, Java rules enforce accepted best practices and coding styles, uncover design issues, detect constructs that are either broken, extremely confusing, or prone to runtime errors, flag code documentation issues, suboptimal code, potential security flaws, and issues when dealing with multiple threads of execution.\\

SpotBugs\footnote{https://spotbugs.github.io/} (formerly FindBugs) is a static analyzer mostly focused on defect detection based on a predefined set of more than 400 bug patterns in Java code. These bug patterns check for bad practices, correctness, performance, malicious code, and security issues. \\

Fortify\footnote{https://www.microfocus.com/en-us/cyberres/application-security/static-code-analyzer} is one of the most popular static analyzers specifically for violations of security-specific coding rules and guidelines in multiple languages. Fortify comprises eight vulnerability analyzers: buffer, configuration, content, control flow, dataflow, null pointer, semantic, and structural.

\subsubsection{Dynamic Program Analysis}
\R{Dynamic program analysis involves analyzing the properties of a program while it is executing with real input data. Unlike static analysis, dynamic analysis aims to identify issues while the code runs. Dynamic analysis is useful in identifying (i) a lack of code coverage, (ii) memory allocation and leaks, (iii) performance bottlenecks, (iv) software vulnerabilities and defects, and (v) concurrency issues such as deadlocks and race conditions.}\\

\R{A comprehensive list of dynamic analysis tools can be obtained from the \url{analysis-tools.dev} website\footnote{https://github.com/analysis-tools-dev/dynamic-analysis}. Some notable \textbf{dynamic analysis tools} include:}\\ 

\R{Valgrind\footnote{https://valgrind.org/} provides a suite of tools for building dynamic analysis tools. For instance, Memcheck can detect many memory-related errors that are common in C and C++ programs and can lead to crashes and unpredictable behavior, Helgrind is a thread debugger that finds data races in multithreaded programs, and Massif performs detailed heap profiling by detecting which parts of the program are responsible for the most memory allocation.}\\

\R{Application Verifier (AppVerifier)\footnote{https://learn.microsoft.com/en-us/windows-hardware/drivers/devtest/application-verifier} by Microsoft is a runtime verification tool for unmanaged code to detect and help debug memory corruptions, critical security vulnerabilities, and limited user account privilege issues that would be difficult to detect during regular application testing. }\\

\R{Code Pulse\footnote{https://code-pulse.com/} is a real-time code coverage tool for penetration testing activities by OWASP and Code Dx (acquired by Synopsys\footnote{https://www.synopsys.com/software-integrity.html}) and automatically detects coverage information while the tests are being conducted.  Code Pulse currently supports Java and .NET Framework programs.}

\begin{tcolorbox}[colback=white!5!white,colframe=blue!75!black]
  \R{\textbf{Teaching Exercises.} A useful exercise for understanding the utility of performing static or dynamic analysis on code is to analyze the impact on code quality before and after a defect-fixing commit. Students will retrieve versions of the code from a specific repository both before and after a defect-fixing commit. By utilizing a chosen analysis tool, such as SpotBugs, students will explore the implications of defect fixes regarding correctness, performance, and security issues.}
\end{tcolorbox}

\subsection{Mined Data Analysis}

The analysis of the collected data depends \B{mainly} on the data type, \B{but also on other factors such as the purpose of the analysis and the tools and techniques available}. Even when source code repositories are analyzed, there are multiple data sources and types: source code, commits, GitHub issues, questions and answers, review comments, code comments, numerical data, etc. 
\B{In this section, we will review some data types and techniques for analyzing the mined data.} 
\R{As we have focused on code in the previous section, we cover a broader range of artifacts in this section. Although source code analysis requires specific techniques, which are beyond the scope of the current chapter, the topic modeling and deep learning-based approaches mentioned in this chapter are widely used for analyzing the source code. }

\subsubsection{Unstructured Data}

``Unstructured data does not have clear, semantically overt, easy-for-a-computer structure. It is the opposite of structured data, the canonical example of which is a relational database, of the sort companies usually use to maintain product inventories and personnel records.'' ~\cite{chen2016survey}

Textual data such as commits \R{or source code} are unstructured, as they do not have a pre-defined format. Instead, the authors can write the text in any order and with different contexts, even though a template is used. Understanding the content of the text is of main interest in many studies and requires analysis. 
We can use several techniques to analyze textual data. The main factor of using the analysis technique is `what' we intend to identify in our study. This can vary from more straightforward techniques, such as analyzing the frequency of words based on Term-Frequency Inverse-Document-Frequency (TF-IDF) metrics, to more in-depth analysis to understand the content of the documents or categorize them, such as topic modeling, clustering, and classification techniques to sophisticated analysis such as finding the reasons, relations, or causality among concepts, using more advanced text analysis techniques such as causal inference. 
\B{In all cases, the textual data could be represented with numerical vectors (also referred to as vector embedding) that can be learned using deep neural models. }

\paragraph{Topic Modeling}
Topic modeling or latent topic modeling is a category of techniques for automatically extracting topics from a corpus of text documents. The corpus of documents can be short or long texts, such as commit messages, app review feedback, or bug reports. Please note that each commit, app review, or bug report is considered a document in this example. A topic is a collection of terms that frequently occur in the documents. 
When topic modeling is applied, the main topics of the documents' corpus are extracted; therefore, we can compute the topics of each document as well. 
Thus, topic modeling uncovers the latent semantic relationships of the documents. As this ensures a faster analysis of many documents, topic modeling is used for several software engineering applications such as bug triaging, finding the most discussed topics or issues, traceability link recovery, and concept location. 
\B{Multiple applications of topic modeling in software engineering research are discussed in the literature~\cite{silva2021topic}.}

Examples of topic modeling algorithms are Latent Semantic Indexing (LSI), Probabilistic Latent Semantic Indexing (PLSI), Latent Dirichlet Allocation (LDA) and its variations, such as Hierarchical Topic Models (HLDA), non-negative matrix factorization (NMF), and Biterm Topic Model (BTM). While LDA and other algorithms are designed for longer text, BTM is a newer algorithm for short text. 
The BTM algorithm considers the Biterms in the whole corpus to enhance the topic learning in short texts. 
Online algorithms for BTM, i.e., online BTM (oBTM) and incremental BTM (iBTM), are also introduced to speed up the inference of BTM on large data sets. 

The algorithms mentioned above are mainly used to extract the topics from documents at a static point in time. 
As the distribution of the terms in a document changes over time, so can its topical topic modeling techniques, which are developed to detect the evolution or variations of topics in time-stamped documents. Dynamic Topic Model is one of these algorithms. 
Online Latent Dirichlet Allocation (OLDA) is another method
that tracks the variations of topics over text streams. The OLDA models the texts' topics of one time slice based on the topics of the last time slice. 
Newer algorithms are developed that consider previous time slices of the documents for better topic modeling. Adaptively Online Latent Dirichlet Allocation (AOLDA) is an algorithm specifically developed for software engineering and app review analysis. It improves OLDA by adaptively combining the topic distributions of previous versions to extract topics in the current time slice. 
Adaptive Online Biterm Topic Modeling (AOBTM) is a similar algorithm developed for short texts. It analyzes the statistical data of previous time slices to identify the topic distribution of the current time slice. 

The more recent topic modeling algorithms are based on word embeddings or neural network-based topic models, such as Top2Vec, LDA2Vec, and BERTopic. 

\textit{Considerations.}
When applying topic modeling, several considerations are necessary to improve the results: 

\begin{itemize}
    \item Text pre-processing: Though the text pre-processing varies based on the study context and data, the common techniques are to remove punctuation and non-textual symbols, check the language of the text (e.g., English only), spell check and correct the spellings of the words, stemming and lemmatization, and removing stop words. 

    \item Investigating the length of the text: It is shown that the topic modeling algorithms do not work correctly for short text. So, investigating the length of the documents is necessary in selecting the topic modeling algorithms, whether they are designed for short or long text. 

    \item The evaluation metrics: The evaluation metrics of topic models span over quality, interpretability, stability, diversity, efficiency, and flexibility. Two well-known terms to measure the topic models are perplexity and coherence. The former refers to how well the model explains the data (predictive power of the model), and the latter identifies a measure of whether the generated words for a topic can be associated with a single semantic meaning. One way to evaluate coherence is Pointwise Mutual Information (PMI). For other evaluation metrics, the work of Abdelrazek et al.~\cite {ABDELRAZEK2023102131} provides a good starting point. 

    \item Number of topics: The user should choose the number of topics in most algorithms. Rather than using a random number or based on the heuristics, a common way is to explore a range of different topic numbers and evaluate the coherence score of the results, which could be a good determination to select the number of topics. There are other metrics for models such as BERTopic that should be evaluated for selecting the best number of topics.

    \item Hyperparameter setting: Similar to the previous point, setting the hyperparameters is important in running the algorithms. So, care should be taken to experiment with different choices before applying the topic modeling. 

    \item Running time and size of data: One main consideration in choosing the topic modeling algorithm is the time it takes to execute the algorithm and whether it can be applied to a large dataset.  
\end{itemize}

\R{\paragraph{Sentiment Analysis.}}
\R{Sentiment analysis, also referred to as opinion mining, refers to gathering and analyzing people's opinions, emotions, or attitudes towards an entity (i.e., individuals, product, topic, etc.)~\cite{wankhade2022survey}. In software engineering, sentiment analysis can analyze various MSR artifacts, such as app reviews, users' product feedback, and developers' discussions. Sentiment analysis determines whether the opinion is positive, negative, or neutral. The analysis can be done at different levels of granularity: document, sentence, phrase, or even certain aspects of an entity. Several algorithms are developed for sentiment analysis, from lexicon-based approaches to unsupervised and supervised machine learning techniques, including deep learning models and pre-trained language models, hybrid approaches, and transfer learning approaches.}

\textit{Considerations.}
While for some text analysis applications, the text should be cleaned and punctuation marks removed, the text pre-processing considerations for sentiment analysis should be done carefully. Several essential features can reveal sentiments, including punctuation marks, emojis, and slang words, which are some selected features for sentiment analysis.\\

\paragraph{\R{(Deep)} Neural Network Based.}
Another common technique for analyzing textual data is representing the document with a vector. This vector, called embedding, can be a non-contextual or contextual text representation. 
For non-contextual representations, \B{the model} learns a fixed vector for each word, no matter their context.

Famous examples of such embeddings are word2vec, glove, and code2vec.
In the contextual representation, the word embedding depends on its surrounding words.
Contextual representation has recently been used in many applications in which the embedding of the words or text is extracted from a language model, a model that learns a probabilistic model of a natural language from a large corpus of textual data. 
Multiple models can be used to extract the embeddings of text or code, including Sentence Transformers \B{or GPT family} for text or code-specific language models such as CodeBERT, GraphCodeBERT, CodeT5, \B{or Code Llama.}
\B{Many newer models can also be used, and we refrain from naming them due to the increasingly fast introduction of newer models. 
While topic modeling and sentiment analysis are techniques aiming at a specific goal (i.e., understanding what is being discussed and what emotion is expressed), deep learning models are generally discussed and can be used for various purposes, including topic modeling and sentiment analysis. These models or embeddings extracted from them can be applied in other analyses such as predictive analysis or training machine learning models. Some examples of the studies that use deep learning for mining software repositories are mentioned in the survey by Yang et al.~\cite{yang2022survey}.}

\textit{Considerations.} 
In the following, we list a few points to consider when choosing the models. 

\begin{itemize}
    \item Cost of using a model. Most of the recent language models are based on deep learning approaches. The computational costs available to the researchers should be considered when using them. Additionally, if using models such as a GPT-based family, one may intend to pay for the calls to the model and receive the embeddings. Therefore, the budget associated with the research based on the dataset size is an important factor.
    \item Model selection. Several non-contextual and contextual models are specifically developed for software engineering tasks and code. Depending on the application, we recommend a search to identify the model to use in the analysis. Several SE-specific language models have been developed, such as seBERT, which is a BERT model trained from scratch on software engineering data. These specific models are available for various applications and domains in software engineering. 
    It is necessary to conduct at least some initial \B{studies} if there is insufficient literature to ensure the final model is chosen properly.

\end{itemize}

\subsubsection{Structured Data}
Structured data is data that has a standard format.
Examples of structured data are time stamps, number of commits, and years of experience. 
The structured data are analyzed through various means, such as correlation analysis, time series analysis, statistical analysis, clustering, regression analysis, and developing prediction models. 
Note that some techniques, such as clustering, \B{and developing machine and deep learning models} might apply to the unstructured data. 
\R{A discussion about machine learning strategies for MSR can be found in~\cite{guemes2018emerging}.}
\B{Both structured and unstructured data can be used for the analysis or extraction of quantitative metrics while working with unstructured data. For example, the mean number of accepted answers on Stack Overflow can be calculated while also analyzing the contents of the questions and answers.}

In the following, we briefly describe some of the statistical tests. 
Statistical methods, or models, are powerful tools for analyzing data and supporting arguments. They are mathematical formulas used to analyze numerical raw data. 
The study conducted by De Oliveira Neto et al. \cite{de2019evolution} provides a survey on different tests that are used for empirical studies in software engineering, which can be used as a starting point to review the widely used statistics tests. 
Parametric and non-parametric tests are specific classes that refer to the distribution of the population's parameters. 
The t-test, ANOVA, and F-test are some common parametric tests. 
Examples of non-parametric tests are Mann–Whitney U test and its variations, Kruskal–Wallis test, and the $\chi ^2$ test. 
If multiple tests are conducted, correction tests such as Bonferroni or its variations should be considered. 

Another method is statistical power analysis is a test conducted to accept or reject a hypothesis. The statistical power refers to the probability of a hypothesis finding an effect if there is an effect. Power analysis uses a significance level, effect size, and statistical power to estimate the required minimum sample size.

\textit{Considerations.}
In the following, we list a few points to consider when analyzing structured data.

\begin{itemize}
    \item Data distribution. When applying statistical tests, it is imperative to assess data distribution. Statistical tests are often developed for normal or non-Gaussian distributions, and one cannot be applied to the other. Therefore, the results are unreliable if the data follows a non-Gaussian distribution, but the test is for normal distributions. Shapiro–Wilk and Kolmogorov–Smirnov tests are commonly used to assess the normality distribution of datasets.

    \item Reporting effect size. When applying statistical tests, it is necessary to understand and reflect on the effect size and significance level and not choose them arbitrarily. When reporting the results of a statistical significance test, usually p-values are reported but not the effect size. However, a p-value only informs the existence of an effect but not its magnitude (i.e., effect size). Therefore, the effect size should be reported to find out the size of a significance~\cite{sullivan2021using}. 
    
    \item Often, to develop an analysis, the model's features combine structured and unstructured data. In this case, the embeddings of the text or code could be used as features, or the text itself could be considered. There are several approaches to be used, and one can experiment with different combinations to evaluate the results. 
    
\end{itemize}

\begin{tcolorbox}[colback=white!5!white,colframe=blue!75!black]
  \R{\textbf{A Do for Educators.} Data analysis depends on the data types. Even for source code repositories, there are multiple data sources and types. Tools must be selected considering the nature of the repositories and the analysis to conduct.}
\end{tcolorbox}

\begin{tcolorbox}[colback=white!5!white,colframe=blue!75!black]
  \R{\textbf{Teaching Exercises.} A useful exercise for understanding data analysis in MSR involves performing topic modeling on defect-fixing and non-defect-inducing commits. Students will start with a set of commits from Kamei et al.~\cite{Kamei2013JITQA} and preprocess the commit messages to ensure data quality. Students will identify the main topics discussed in defect-fixing and non-defect commits using a specified topic modeling technique, such as Latent Dirichlet Allocation (LDA) or BERTopic. They will analyze and interpret these topics, comparing the topics of defective and non-defective commits and exploring their relevance to different types of defects and their implications for software maintenance practices. A similar practice can be applied using sentiment analysis to investigate whether the sentiments of commits change when fixing a defect compared to a non-defect one.  
  }
\end{tcolorbox}

\subsection{Data Synthesis}

\R{Collecting and synthesizing the analysis results is crucial to interpreting data and discovering important findings.}
Before presenting all the results extracted from data analysis, an MSR researcher's primary focus is to remember the goal addressed in the previous steps.
Following the GQM approach (presented in Section \ref{msr_process:gqm}), all the plots and the results shown in the study aim to answer the research questions formulated.
Therefore, before starting, it is crucial to keep the following points in mind:
\begin{itemize}
    \item \textbf{Research Objectives as a Guide:} Using the formulated research goals helps to organize and prioritize your results. Therefore, considering a mapping link between the research question and the findings helps to find the best solutions to present results through discussion, plots, and tables.
    \item \textbf{Relevance to Research Questions:} Articulating how each result addresses specific research questions helps to check if the goal is accomplished. This approach ensures that the presentation of the results maintains a cohesive narrative and directly ties back to the study's purpose.
    \item \R{\textbf{Embracing Unexpected Findings:} During an MSR study, it is possible that the results show something different from what is expected. The presentation and discussion of unexpected results could enlighten new implications and open the possibility of novel opportunities.}
\end{itemize}

Once the goal is clearly assessed, it is necessary to format the data and the outcome of the analysis to obtain the information needed to answer our research questions.
Therefore, summarizing and exploring the data is important to extract the veiled message from large collected data.

The main approaches to summarize the analysis results are based on descriptive metrics and visual representations.

\subsubsection{Descriptive Metrics}
Using properly descriptive metrics and 
presenting information is integral to conveying the study results.
In summarizing metrics derived from MSR, descriptive statistics are pivotal in distilling complex datasets into meaningful and interpretable insights.
The choice of descriptive statistics should be adopted, focusing on the goals of the analysis to find the most suitable answers from the data. 
Key descriptive statistics capture the central tendency, variability, and distribution of the metrics, including mean, median, mode, and standard deviation~\cite{kitchenham2017robust}.
For instance, the mean serves as a central measure, offering an average value that summarizes the overall trend within a dataset. \R{However, it is important to note that the mean can be influenced by extreme values, making it less representative of skewed distributions. In the context of software components, calculating the mean number of lines of code provides an overview of the project's codebase size, but it may not fully capture the central tendency if the distribution is highly skewed.}
The median, resistant to extreme values, provides a robust representation of the center of the data.
In MSR, identifying the median response time for resolving issues can offer a more stable representation of the typical resolution time, even in the presence of outliers.
Mode highlights the most frequently occurring values, emphasizing prevalent patterns.
For instance, identifying the mode in a dataset of commit frequencies may highlight the most common development activity intervals.
Standard deviation quantifies the dispersion of data points, offering insights into the dataset's variability.
For example, examining the standard deviation of code churn rates can reveal the extent to which development activity is variable across different periods.
Collectively, the combination of several descriptive statistics contributes to a comprehensive understanding of the metrics, aiding researchers and stakeholders in uncovering patterns, trends, and central characteristics within the intricate landscape of software development data.

\subsubsection{Visual Representations}
Descriptive statistics offer a concise and straightforward way to present data analysis outcomes.
However, relying solely on numerical summaries, such as means and medians, can obscure significant patterns and variations within data.
This approach may also fail to capture the complexity of relationships or variations in multidimensional data.
To overcome these shortcomings and improve the interpretability of the results, it is crucial to supplement numerical summaries with visual representations. Visualizations provide an interactive and intuitive method for discovering patterns, outliers, and correlations within the data. Employing various visual elements such as plots, tables, and representative graphs when synthesizing your data will increase the comprehension of the readers interested in the study and allow them to catch it.

As for the descriptive metrics, each plot type serves a unique purpose in conveying information effectively.
Some of the most used visual representations to summarize the results in MSR studies are:

\begin{itemize}
    \item \textbf{Line Charts}: This type of plot is helpful to visualize temporal aspects of software repositories, showcasing the evolution or the history of metrics over time. An example of this type of plot for MSR Studies can be related to analyzing the use of third-party libraries over time~\cite{Salza2018Third-PartyLibraries}.

    \item \textbf{Bar plots}: These plots show the frequency of a categorical variable using bars. The bars' height indicates the data's value in each category, such as the frequency, total count, sum, or average. An example of using bar plots for MSR studies is to analyze the number of occurrences of bug categories in software systems~\cite{Catolino2019Bugs}.
    
    \item \textbf{Box Plots}: These plots show the distribution of a numerical variable using five statistics: the minimum, the lower quartile, the median, the upper quartile, and the maximum. Box plots can be used to compare the central tendency and the variability of different groups or samples of data. An example of using box plots for MSR studies can be related to the extraction of the number of commits that lead to the appearance of a code smell~\cite{Tufano2015WhenAndWhy}.
    
    \item \textbf{Scatter Plots}: This type of plot is used to show the relationship between two numerical variables. The position of each dot on the horizontal and vertical axis indicates the values of the variables for each observation or unit of analysis. Scatter plots help explore the correlation or association between two variables or identify outliers or clusters in the data. Scatter plots are commonly used to analyze a specific factor's evolution to find tendencies. An example of using scatter plots is analyzing the adoption of reusability mechanisms in source code over time to find common patterns between projects~\cite{giordano2024adoption}. 
    
    \item \textbf{Network Graphs}: This type of plot is commonly used to model relationships between elements. It can be used to visualize clusters and patterns in the data and analyze complex systems' properties and behaviors. Network graphs in MSR studies are commonly used to explore collaborative networks between developers~\cite{Gupta2014OrgMSR}.
    
\end{itemize}

\subsubsection{\R{Best Practices and Dos for Educators}}

The following \R{best practices} are collected to inform MSR researchers of the main requirements for reporting an MSR study. 

\paragraph{Replicability}
In MSR studies, it is crucial to document data sources, collection methods, preprocessing steps, and analysis techniques for easy replication by other researchers. MSR researchers should always provide all the material to reproduce the conducted studies, including code, scripts, or workflows. All this content should be in an accessible collection called \textit{replication package}.
Mahmood et al.~\cite{MAHMOOD2018ReproducibilityDP} proposed using common online replication services, such as OpenML\footnote{http://www.openml.org/} and Zenodo\footnote{https://zenodo.org/}, to replicate MSR studies effectively.

\paragraph{Use of Visual Elements} When incorporating visual elements in information representation, prioritize clarity, consistency, accessibility, and relevance.
Visualizations should be designed with simplicity to ensure easy comprehension for a diverse audience. Consistent use of visual elements fosters a cohesive visual language, while accessibility features, such as color-blind-friendly palettes and alternative representations, promote inclusivity.
Additionally, colors can be strategically used to represent additional information, adding depth to the visualization. Visualizations should serve a clear purpose and contribute meaningfully to the overall narrative, enhancing the interpretability of complex data.

\paragraph{Use of Relative Discussions to Findings} When presenting research findings, it is important to emphasize their significance and relevance to the research questions.
Researchers should discuss existing literature, industry benchmarks, or expectations to comprehensively understand the relative importance of the findings.
By highlighting the connections between the findings and the external contexts discussed, researchers can offer richer insights and demonstrate the practical implications of their work. This approach ensures that the reported results are not isolated but instead contribute meaningfully to the ongoing goal of the study.

\paragraph{Use of a Systematic Quality Assessment Process}
When all research components are in place, employing a systematic quality assessment process is prudent. This systematic approach ensures a thorough and rigorous evaluation of all artifacts generated throughout the study. The assessment process should encompass a comprehensive review of the dataset, code, visualizations, and other materials, verifying their accuracy, completeness, and adherence to established standards.
MSR researchers can identify and rectify potential errors or inconsistencies before finalizing the study's outcomes.
To have an effective systematic evaluation process of the MSR study, Chatterjee et al.~\cite{Chatterjee2022EmpiricalStandards} proposed a series of standards that an MSR researcher should use to ensure high-quality artifacts.
\begin{tcolorbox}[colback=white!5!white,colframe=blue!75!black]
  \R{\textbf{Teaching Exercises.} 
   A good exercise is reproducing an existing \B{study}, including related synthesis, analysis, and interpretation of the results. Considering the outcome of the previous exercise that highlights the most used topics in defect-fixing commits, students will attempt to synthesize the data, collecting key elements and interesting findings. They will create plots to observe critical characteristics of the synthesized data and make interpretations of the results to explore the topics in defect-fixing commits.}
\end{tcolorbox}

\subsection{Threats to Validity in MSR Studies}
While often considered a mere afterthought of mining processes~\cite{verdecchia2023threats}, Threats to Validity (TTV) play an essential role in empirical inquiries. Summarily, threats to validity could potentially affect the accuracy or credibility of a study or its results~\cite{wohlin2012experimentation}. A transparent, comprehensive, and truthful documentation of the TTVs that may have influenced a study is essential. TTV considerations are crucial to let the reader accurately understand the mining process, interpret its results, and potentially build upon them. 
TTVs and related mitigation strategies should be considered throughout the entirety of the mining processes, starting from the earliest stages (e.g., by reflecting if the RQ is correctly formulated to achieve the research goal) till the concluding steps.  
Among different TTV categorizations, often MSR studies follow the four categories presented by Wohlin et al.~\cite{wohlin2012experimentation}, namely \textit{conclusion validity}, \textit{internal validity}, \textit{construct validity}, and \textit{external validity}. 

\subsubsection{Conclusion Validity}
Threats to the conclusion validity refer to impediments that may affect the ability to draw the correct conclusion about relations in the collected data. Leaving aside conclusion threats related to statistical result analyses (e.g., low statistical power and violated statistical test assumptions), recurrent conclusion threats in MSR processes regard confounding factors, such as measure reliability (mainly if dynamic code analyses are used), random irrelevancies in the mining process (e.g., including extended periods of repository inactivity), and excessive heterogeneity of repositories (e.g., considering in the same MSR process both cloud-native and embedded contexts). Mitigation strategies for conclusion TTVs in MSR studies often entail the selection of repositories based on \textit{a priori} defined criteria, the carefully motivated use of sound source code analysis tools, and \textit{post hoc} scrutiny of results trends to spot potential conclusion pitfalls related to specific repositories.  

\subsubsection{Internal Validity}
Threats to the internal validity regard unknown factors that can impact the study's relationship between the phenomenon considered and the observed results. Threats of this nature are often related in MSR processes with a summary repository selection, inaccurate mined data post-processing, unfitted statistical data analyses, and, when dynamic analyses are used, the influence of previous code executions on subsequent ones. Internal TTVs of MSR studies can often be tackled during the mining design phase by ensuring that a systematic, repeatable, and documented process is used to select repositories and manage the collected data, and ``cooldown'' and cache cleaning precautions are taken to start each dynamic measurement as a clean slate.

\subsubsection{Construct Validity}
Construct validity pertains to the representativeness of the designed mining process to accurately study the considered theoretical construct. Most commonly, construct threats in MSR processes relate to the design phase of the mining processes and may involve a vague or ill-suited definition of the construct (e.g., defining technical debt as maintainability issues), under-representation of the considered construct (e.g., considering a single type of test flakiness to study the topic), or the adoption of an overly-narrow analysis to examine a considerably broader construct (e.g., focusing solely on number of functions to study software testability). During the design of MSR processes, the construct under study and how the mining considers such construct should be carefully discussed among miners to understand, mitigate, and document the potential construct TTVs entailed by the adopted methodology. 

\subsubsection{External Validity}
External validity regards the extent to which the results obtained with the mining process can be translated into other contexts (e.g., industrial practice, another application domain, or a different programming language).

Among the external TTVs, one of the most common ones in MSR regards the under-representation of the entire set of software repositories and commits relevant to answering the RQs. A strategy to systematically sample repositories and commits is paramount to mitigate this threat by considering the threat such sampling entails. A sound sampling should carefully select both a fitting sampling technique (e.g., stratified sampling if repositories of different natures need to be represented) and a sample size. Regarding sample size, numerous tools can be used to calculate it based on confidence interval and margin of error, typically 95\% and 5\%, respectively, but these can vary according to the entire set of repositories considered.

Another common external TTV of MSR studies is adopting \textit{ad hoc}, manual, or outdated processes to mine and analyze the source code. Conscious attention should be paid to ensure that systematic and repeatable state-of-the-art and practice processes are used, i.e., analysis tools and considered repositories, to mitigate this threat.

Additionally, depending on the data under study, other sampling techniques, such as stratified sampling, should be considered to ensure that the distribution of the sample data in each class or group is retained and similar to the distribution of the classes in the original data.

\section{\B{Complementing Software Repository Mining Studies}}
MSR studies might be combined by analyzing additional sources of information. From a methodological standpoint, this is the basis of \emph{mixed-method} research~\cite{creswell1999mixed,storey2024guidelines}, a research approach combining qualitative and quantitative research methods elements within a single study or research program. This approach seeks to harness the strengths of both methodologies to provide a more comprehensive and nuanced understanding of a research question or phenomenon. \B{Storey et al. \cite{storey2024guidelines} recently defined guidelines aiming to ease the application of mixed-method research in software engineering, as well as a catalog of best and bad practices to help apply it. In the scope of MSR research, mixed-method research may be useful to enhance or confirm the findings coming from mining human-generated data but also enable asking different questions as they arise during the study \cite{storey2024guidelines}. In doing so, researchers may triangulate the findings by exploiting multiple research methods, increase the overall credibility of the findings, or even find contradictory or surprising results \cite{storey2024guidelines}.}

\B{In the following section, we overview the potential methodologies that may be used to complement the results of MSR studies and provide exemplary articles that may be used to illustrate these methodologies in practice.}

\smallskip
\textbf{\B{Complementing MSR Research with Qualitative Methods.}} For instance, imagine complementing an MSR exploration with qualitative research methods, such as \emph{surveys}, \emph{interview studies}, and \emph{focus groups}. A survey represents a research method that involves collecting data from a sample of individuals by administering a structured set of questions. When combined with MSR studies, a survey could capture the subjective experiences and perceptions of developers, project managers, and other stakeholders, providing a human narrative to complement the quantitative trends identified in the MSR exploration. Literature has often relied on this combination of research methods. A notable example is the paper by Qiu et al.~\cite{qiu2019going}, where the authors effectively combined software repository analyses and insights from a survey study to investigate the impact of social capital on the sustainability of open-source projects.

Interviews represent an alternative to survey studies. An interview is a qualitative research method that systematically collects and analyzes data from one-on-one interviews with participants. Unlike survey studies, interviews provide researchers with finer-grained insights from the interviewees' experiences. By nature, interview studies can only reach a small sample size and are typically limited to the analysis of a few practitioners. As a consequence, interview studies are particularly suitable when the aim is to gain a deep and nuanced understanding of participants' experiences, perspectives, or beliefs. For example, interview studies with key stakeholders offer a deeper understanding of individual experiences, motivations, and decision-making processes, adding a layer of context to the automated data. On the contrary, if the goal is generalizability, there might be better research instruments than an interview study. The interested reader might take the paper by Tao et al.~\cite{tao2012software} as a valuable example of how to make interviews instrumental for the goals of an MSR exploration. 

Focus groups represent an additional alternative. These refer to a qualitative research method that involves a small, diverse group of participants who engage in an open and facilitated discussion about a specific topic under the guidance of a moderator. This method aims to gather insights, perceptions, opinions, and attitudes through group interaction, allowing participants to express their views and respond to each other in a dynamic setting. Like the interview studies, focus groups rarely have the power to generalize the insights that emerge and should be used to understand better the quantitative findings obtained by an MSR study. An example of applying this combination can be found in the paper by Falessi et al.~\cite{falessi2018empirical}.

\smallskip
\textbf{\B{Complementing MSR Research with Quantitative Methods.}} Besides qualitative research methods, it is also worth reporting that an MSR study can be empowered using additional mining instruments beyond the scope of the traditional version control systems. For instance, \emph{issue trackers}, where bugs and tasks are recorded and discussed, become a valuable trove of information. Mining these repositories unveils the challenges developers face, the evolution of bug resolution processes, and the collaborative dynamics surrounding issue resolution. At the same time, \emph{code review repositories}, theaters where code changes are scrutinized, are an opportunity to explore the quality assurance practices within a project. By analyzing discussions, comments, and decisions made during code reviews, researchers gain insights into coding standards, knowledge transfer, and the social dimensions of code evaluation. Finally, \emph{developer forums} like Stack Overflow become digital arenas where practitioners seek and provide solutions. Mining these forums provides a glimpse into the knowledge-sharing ecosystem, exposing common challenges developers face and the collaborative solutions the community offers. A notable example of the combination of multiple mining instruments can be found in the work by Ram et al.~\cite{ram2018makes}.

\smallskip
\textbf{\B{Key Advantages of Mixed-Method Research.}} 
In the scope of MSR, integrating qualitative research methods and exploring additional mining instruments offer a nuanced perspective. While MSR studies provide the backbone of empirical evidence, qualitative methods infuse a \emph{human dimension}, unraveling the stories behind the code changes. Delving into issue trackers, code review repositories, and developer forums adds layers of context, portraying software development as \emph{a collaborative, evolving journey rather than a mere compilation of code changes}. In essence, this integration transcends the boundaries of quantitative and qualitative methodologies, fostering a research approach that mirrors the complexity of the software development ecosystem. It is an opportunity to see automated analyses and human narratives converging. It offers researchers a holistic understanding of the intricate connection between code, collaboration, and the people who bring software projects to life. In Appendix \ref{sec:repoList}, we report a non-exhaustive list of repositories that the reader may find helpful to running mixed-method MSR research.

\begin{tcolorbox}[colback=white!5!white,colframe=blue!75!black]
  \B{\textbf{A Do for Educators.} MSR studies can be combined by leveraging other methods to obtain mixed-method research.}
\end{tcolorbox}

\section{Ethical Mining}
\B{MSR research typically involves analyzing human-generated data, including developers' activities and interactions in repositories, as well as cultural and geographic data, such as racial and ethnic origin, which are necessary for studies on the geo-cultural dispersion of software communities. Although it is classified as a data-driven strategy rather than a respondent-driven one \cite{storey2020software}, ethical considerations may still be relevant. For example, consider the case of a mining study aiming to investigate PR acceptance rates in open-source repositories. The study may require ranking contributors based on their PR acceptance rate: while GitHub data may indeed be used to rank contributors and identify the most/least successful contributors, publishing the identity of those contributors would be unethical.} 

Despite ethics plays a role in MSR research, it is often overlooked. From 100+ papers on MSR mining challenges and data showcases from 2006 to 2021, only a few discussed ethics or data anonymization as part of the threats to validity~\cite{gold2022ethics}, hence suggesting the need for educating the next generation of researchers to consider ethical aspects while mining software repository data. \B{This is especially true when considering the potential impact of MSR research in practice: according to Feitelson \cite{feitelson2023we}, open-source developers are indeed largely open to research, provided it is done transparently.}

Considering source code repositories as an example, it is essential to note that publishing source code under a license differs from publicly releasing a repository. Repository data are often not explicitly licensed for study and unrestricted use. Therefore, ethical concerns might arise, and ethical issues should be considered in such situations~\cite{gold2022ethics}.
\B{More particularly, when mining the human-generated data coming from repositories, some recommendations based on the Menlo report include~\cite{gold2022ethics}:}

\begin{itemize}
    \item \emph{Stakeholder identification} - Consultation of all parties involved and impacted by the research (e.g., ICT researchers, human subjects, non-subjects, users, and platform owners) before using their data for research purposes. \B{Adhering to this guideline when performing large-scale mining analyses may be challenging. Nonetheless, it is worth remarking that, according to the GitHub's acceptable use policy,\footnote{The GitHub's acceptable use policy: \url{https://docs.github.com/en/site-policy/acceptable-use-policies/github-acceptable-use-policies}.} researchers \emph{``may use public, non-personal information from the Service for research purposes, only if any publications resulting from that research are open access''}. This applies to both data directly extracted by GitHub, e.g., through GitHub's APIs, and data indirectly coming from GitHub, e.g., publicly available datasets that derive from GitHub data. As such, one approach is to focus on using non-personal data and ensuring that research findings are published as open access. Additionally, researchers may engage with repository maintainers and community members through public communication channels to inform them about the research and invite feedback. On the one hand, this may help ensure research transparency. On the other hand, this may foster a collaborative environment, even if direct consultation with every individual contributor is impractical.}

    \smallskip
    \item \emph{Informed consent} - ensuring that permission is sought, participation is voluntary, and data is anonymized, carefully processed, stored, and discarded are all essential aspects to be considered. \B{When analyzing large projects with numerous contributors, obtaining informed consent from all participants can be impractical because many contributors might have left or lost interest in the project. In such cases, ethical considerations include respecting the contributors' privacy and ensuring that their data is used responsibly. On the one hand, GitHub's acceptable use policy solely allows the use of public, non-personal information for research purposes. On the other hand, one approach is to anonymize or use pseudonyms for contributors' names, as this protects their identities while still allowing for meaningful research.}

    \smallskip
    \item \emph{Risks and benefits balancing} - the need to consider the potential harms, personal data protection, and the impact of the results.

    \smallskip
    \item \emph{Fairness and equity} - the need to consider the fair selection of subjects, data availability, and fair treatment of parties involved in the study.

    \smallskip
    \item \emph{Compliance, transparency, and accountability} - legal compliance should be considered as part of data handling. For instance, dealing with personal data may require the researcher to comply with the law of the country they are conducting the research. 
\end{itemize}

Some points to consider when performing MSR studies include:

\begin{itemize}
    \item A change of mindset from \emph{``Here is a dataset, let us see what we can find''}, as this can be risky for the participants. So, the ethics considerations should assess the potential areas of harm, including the observations and judgments, and evaluate the impact the research results can have on the individuals/developers. 

    \smallskip
    \item MSR studies are often conducted on a sample of the data, which is random. Ethics could play a role in ensuring the sample represents the people involved and is inclusive regarding the questions and the process. The ethical consideration ensures that individuals and analyses are not excluded from the results.

    \smallskip
    \item There are different laws and regulations regarding privacy, including the EU's General Data Protection Regulation (GDPR) and the California Consumer Privacy Act of 2018 (CCPA), in addition to the Intellectual Property (IP) laws. Various licenses also state different usage allowances. These laws sometimes restrict specific usages and research and should be considered when designing the study or before data collection. However, for the sake of simplicity, restricting research repositories with licenses that permit studies could affect the generalizability of the research findings. For instance, this aspect should be considered and actively discussed in the paper as a threat to validity. 
\end{itemize}

\B{In conclusion, educators should proactively integrate ethical MSR practices in their courses. First and foremost, it is paramount to highlight the significance of adhering to GitHub's acceptable use policy and safeguarding contributors' identities by refraining from disclosure without explicit consent. Additionally, educators can foster student comprehension of the legal and regulatory landscape governing data privacy and intellectual property rights through interactive discussions and hands-on learning experiences. Employing awareness-targeting teaching methods like quizzes and serious games may be an effective strategy to stimulate critical thinking on ethical challenges inherent in MSR research. An exemplary illustration of this pedagogical approach was provided by Teo et al. \cite{teo2023would}, who introduced an interactive, scenario-based ethical AI quiz for students to self-assess their awareness and perceptions regarding AI ethics.}

\smallskip
\B{\emph{Concrete examples.} To exemplify how ethical concerns may be effectively addressed in the context of MSR research, we briefly discuss the strategies employed by two relevant articles published at the Mining Software Repositories Conference. The first, authored by Yamashita et al. \cite{yamashita2017software}, collected evolutionary data concerned with the programming skills of practitioners to publicly release an open dataset. Before releasing the data, the authors re-wrote the whole change history of the Git repositories so that sensitive information, e.g., developer names and contact details, were removed or changed consistently to protect privacy and confidentiality. The second article, authored by Gonzalez-Barahona et al. \cite{gonzalez2015metricsgrimoire},  released a suite of tools to extract data from software repositories, also contributing a database of human-generated data coming from open-source communities. In their article, the authors explicitly mentioned that the use of data was allowed by the organization providing those data. These two articles represent two valuable examples of how ethics should be preserved while mining software repositories and may be used by educators as case studies.}

\begin{tcolorbox}[colback=white!5!white,colframe=blue!75!black]
  \B{\textbf{A Do for Educators.} MSR studies must be compliant to ethical aspects and regulations enforced by policymakers. Educators should proactively introduce ethical MSR practices in their courses, for instance, by employing awareness-targeting teaching strategies.}
\end{tcolorbox}

\section{\R{Recent Trends and Future Outlook for Educators Leveraging Mining Software Repositories}}
\R{As a last part of this book chapter, let us reflect on the recent trends in MSR research, with an outlook on the future development of the field. 

The most recent advances made by the research community over the last few years reflect the dynamic nature of software development practices and the increasing integration of advanced technologies. One prominent trend is the growing emphasis on leveraging artificial intelligence (AI) techniques to extract insights from vast repositories of software-related data. For instance, the 2024 Mining Challenge Track of the 21$^{th}$ International Conference on Mining Software Repositories (MSR 2024) has featured a challenge on developers' ChatGPT conversations.\footnote{\R{MSR 2024 Mining Challenge: \url{https://2024.msrconf.org/track/msr-2024-mining-challenge?}}} Such an example underscores the relevance of AI-driven approaches in understanding developer interactions, collaboration patterns, and decision-making processes. Educators responsible for mining software repository courses can leverage this trend to provide students with hands-on experience in applying AI techniques to analyze and interpret software-related data, fostering a deeper understanding of the complexities of modern software development processes.

Furthermore, there is a notable shift towards exploring the dynamics of software ecosystems within MSR education. This trend reflects the recognition of software systems as complex socio-technical ecosystems comprising diverse stakeholders, technologies, and dependencies. Educators are called to increasingly incorporate modules on software ecosystems into mining software repository courses, enabling students to gain insights into the interconnectedness of software projects, the evolution of software ecosystems over time, and the impact of ecosystem characteristics on software quality and maintainability. By incorporating software ecosystem analysis into their curriculum, educators can empower students to navigate the intricacies of real-world software development scenarios and equip them with the skills necessary to contribute meaningfully to software projects within diverse ecosystem contexts.

Finally, MSR education is increasingly emphasizing interdisciplinary collaborations and integrating diverse data sources and methodologies. Educators recognize the value of combining traditional MSR techniques with insights from machine learning, natural language processing, and social network analysis to address complex research questions and emerging challenges in software engineering. By fostering interdisciplinary collaboration and exposing students to various methods and tools, educators can prepare them to tackle real-world software engineering problems effectively and drive innovation in mining software repositories.}

\section{Sample List of Repositories}
\label{sec:repoList}

\begin{quote}
    - Bugzilla: https://www.bugzilla.org/\\
    - Bitbucket: https://bitbucket.org/ \\
    - GitLab: https://gitlab.com/ \\
    - Azure Repos: https://azure.microsoft.com/en-us/services/devops/repos/ \\
    - Google Code: https://code.google.com/archive/ \\
    - Jira: https://www.atlassian.com/software/jira \\
    - ProjectLocker: https://www.projectlocker.com/ \\
    - CloudForge: https://cloudforge.com/ \\
    - Zenodo: https://zenodo.org/ \\
    - Awesome GPT: https://gpt4.tools/ \\
    - Docker Hub: https://hub.docker.com/ \\
    - Kaggle: https://www.kaggle.com/ \\

    \textbf{Discussion forums}\\
    - Reverse Engineering: https://reverseengineering.stackexchange.com/ \\
    - Software Engineering: https://softwareengineering.stackexchange.com/ \\
    - Software Quality Assurance and Test: https://sqa.stackexchange.com/ \\
    - GenAI: https://genai.stackexchange.com/ 
    - DevOps: https://devops.stackexchange.com/ \\
    - Hash Node: https://hashnode.com/ \\
    - Dev: https://dev.to/p/editor\_guide \\
    - Code Project: https://www.codeproject.com/ \\ 
    
    \textbf{Online coding platforms used by developers:}\\
    - Code Pen: https://codepen.io/ \\
    - Replit: https://replit.com/ \\
    
\end{quote}

\bibliographystyle{alpha}
\bibliography{references}

\end{document}